\documentclass[twocolumn,showpacs,aps,prl,amsmath,amssymb,superscriptaddress]{revtex4-1}

\usepackage{graphicx}
\usepackage{dcolumn}
\usepackage{bm}
\usepackage[english]{babel}
\usepackage{amsmath}
\usepackage{float}
\usepackage{color}

\usepackage{soul}

\begin{document}

\title{Diffusive and arrested-like dynamics in currency exchange markets}

\author{J. Clara-Rahola}
\affiliation{Department of Applied Physics, University of Almer\'ia, 04120, Almer\'ia, Spain}
\affiliation{i2TiC Multidisciplinary Research Group, Open University of Catalonia, 08035, Barcelona, Spain}
\author{A.M. Puertas}
\email{apuertas@ual.es, jclarar@uoc.edu}
\affiliation{Department of Applied Physics, University of Almer\'ia, 04120, Almer\'ia, Spain}
\author{M.A. S\' anchez-Granero}
\affiliation{Department of Mathematics, University of Almer\'ia, 04120, Spain}
\author{J.E. Trinidad-Segovia}
\affiliation{Department of Economics and Business, University of Almer\'ia, 04120, Spain}
\author{F.J. de las Nieves}
\affiliation{Department of Applied Physics, University of Almer\'ia, 04120, Almer\'ia, Spain}

\date{\today}

\begin{abstract}
This work studies the symmetry between colloidal dynamics and the dynamics of the Euro--US Dollar currency exchange market (EURUSD). We consider the EURUSD price in the time range between 2001 and 2015, where we find significant qualitative symmetry between fluctuation distributions from this market and the ones belonging to colloidal particles in supercooled or arrested states. In particular, we find that models used for arrested physical systems are suitable for describing the EURUSD fluctuation distributions. Whereas the corresponding mean squared price displacement (MSPD) to the EURUSD is diffusive for all years, when focusing in selected time frames within a day, we find a two-step MSPD when the New York Stock Exchange market closes, comparable to the dynamics in supercooled systems. This is corroborated by looking at the price correlation functions and non-Gaussian parameters, and can be described by the theoretical model. We discuss the origin and implications of this analogy.
\end{abstract}

\maketitle
Slow dynamics is common to multiple physical systems such as atoms, granular, and soft-matter systems, as all of them exhibit universal features when approaching the transition towards glass or jammed states \cite{Dove2003, Arimondo2015, Giordano2016, Franklin2015, Josserand2000, Liu2003, Cipelletti2002}. A particular hallmark of such framework is the transition towards fluctuation distributions far from Gaussian, usually characterized with long tails, that depict slow structural relaxations within such arrested systems \cite{deVegvar1993, Castillo2007, Rigbyt1990, Gardel2009, Duri2005, Sessoms2010}. Different models have been proposed to reproduce these observations \cite{Kob2007, March1991}. Within this regard, soft matter systems such as colloids, polymers or surfactants have been established as canonic in non equilibrium systems by their own right \cite{AFN-AMP2016, Jones2002}, also exhibiting strong symmetries with many other fields, for example, with atomic or molecular systems \cite{Pusey1986, Crocker1994, Pham2002}. Such symmetry is originated because these are many-body systems of interacting particles, described by equilibrium and non-equilibrium statistical mechanics \cite{Poon2004}. 

Another field that is strongly amenable to be described by statistical mechanics is financial markets \cite{Meyers2011}, where statistical mechanics has proven to be a useful tool. Louis Bachelier's PhD Thesis, \emph{Theory of Speculation} triggered, a century ago, an increasing interest for finance from a physical and mathematical point of view \cite{Bachelier2006, Fama1972, Fama1976, Bouchaud2000a}. The introduction of computational techniques allowed the development of models such as the fractal one from Mandelbrot, which in fact resembles fractal descriptions also appearing in colloidal structures \cite{Mandelbrot1997, Peters1996}. However, it has been after the works by Stanley et al. \cite{Stanley96} when the amount of research papers published by physicists, in economics in general and in finance in particular, has become relevant. Other achievements such as the GARCH model or the Black-Scholes equation helped in the rise of the field known as econophysics, which aims to employ physical theories and models in finance \cite{BS1973, Mantegna1996, Francq2010, Mantegna2007, Richmond2013}. Still, when considering a physical scope, market dynamics is not fully understood; usually stochastic processes, statistical mechanics or non-linear physics are considered when describing market dynamics, but an unified body that describes the equivalence between mass or lengths with financial magnitudes is lacking.

In this paper we study financial markets, namely foreign exchange markets focused on the Euro-US Dollar exchange rate (EURUSD), from a physical approach, typical of undercooled systems. We study the distribution of the variation of the EURUSD price, and analyze it with a theoretical model of supercooled colloids, where the particles are transiently trapped, but can escape on a large time scale. EURUSD dynamics is considered by computing the mean squared price displacement (MSPD), the analogous to the particle mean squared displacement (MSD). While yearly EURUSD fluctuation distributions can be described by the colloidal glass model, MSPDs are diffusive at all times, without a clear hallmark of glassy physics. Arrested states are however found in particular daily time frames, hallmarked by two steps in the evolution of the MSPD, the corresponding correlation function and also, by the equivalent to the non-Gaussian parameter. The description of the dynamics of colloids and the EURUSD exchange rate, poses the question of a symmetry or possible unification between both fields. Therefore, not only a descriptive approach but a physical origin to foreign exchange markets is proposed, where the EURUSD market, is analog to a supercooled colloidal system.
\begin{figure}
  \includegraphics[scale=0.55]{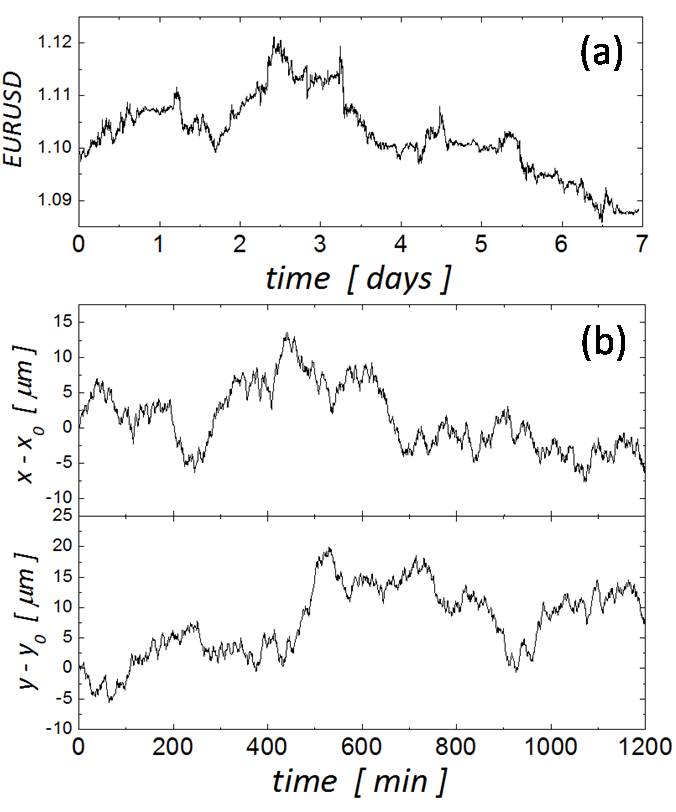}
  \caption{Time evolution of the EURUSD price (a) and cartesian trajectories from a colloid embedded in a quasi 2-dimensional colloidal glass (b).}
  \label{trajectories}
\end{figure}

We have selected the EURUSD instantaneous price at time intervals of 1 minute. The price trajectory strikingly resembles the trajectory of a colloidal particle in suspension, shown in Fig. \ref{trajectories}. Intrigued by this resemblance, we wonder if the dynamics from currency exchange markets and the one from colloids exhibit a more significant symmetry and if ultimately can be described by equivalent physical laws. Note that colloidal dynamics is driven by parameters such as interparticle interactions, density or applied shear \cite{Puertas2002, Puertas2007, Clara2014, Clara2015}, whereas market dynamics is determined by parameters of very different nature, such as trading volume, number of investors at a given period or balance between supply and demand \cite{Kaizoji2002, Duarte2016, Stosic2016}.

\begin{figure}
  \includegraphics[scale=0.63]{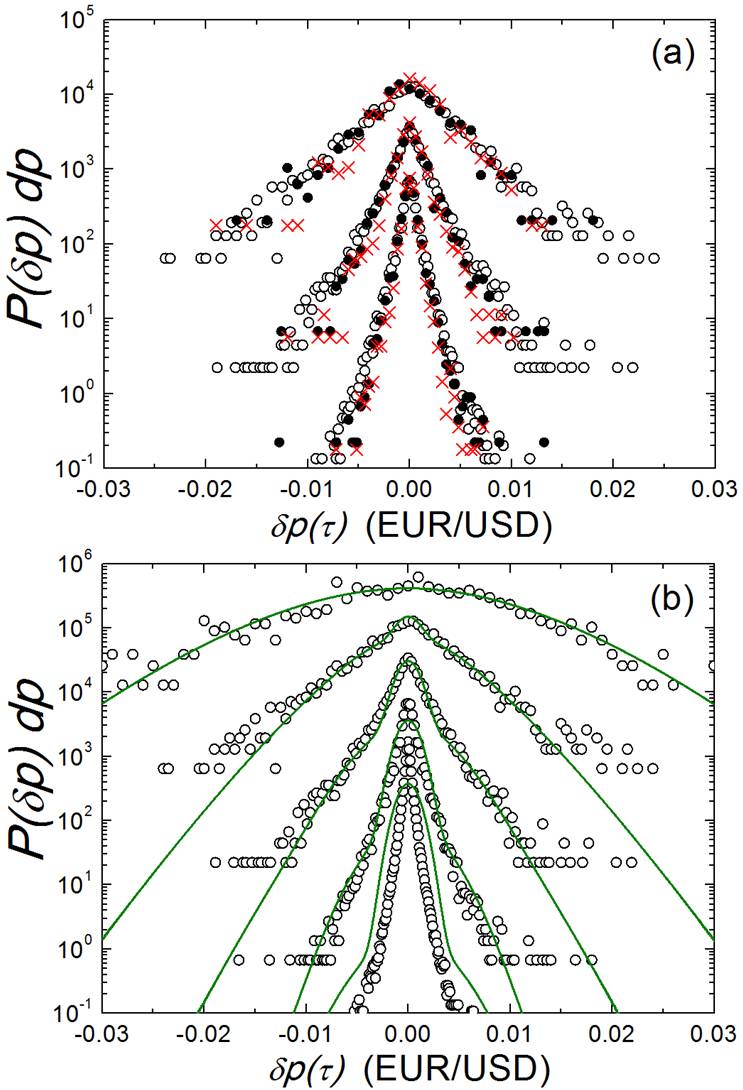}
  \caption{EURUSD fluctuation distributions, $\delta(\tau)$ (a), for $2010-2015$ (open circles), $2001$ (closed circles) and $2007$ (crosses). From bottom to top, $\tau = 25,\:125$ and $625$min. (shifted vertically for clarity). Fit to the experimental pdf (b) over the period $2010-2015$ with the van Hove function of Eqn. \ref{model1} (lines). From bottom to top, $\tau = 5,\:25,\:125,\:625$ and $3125$min. (shifted). 
  \label{pdf}}
\end{figure}

We first study price fluctuations in the EURUSD market, i.e. the difference between prices separated by a lag time $\tau$, $\delta p(\tau) = \langle p(t_{0}+\tau)-p(t_0) \rangle$, where the brackets indicate averaging over different time origins, $t_{0}$ \cite{Note1}. By fixing the magnitude of $\tau$, we construct the probability distribution function (pdf) of $\delta p(\tau)$, an analogue to the standard particle displacement distribution. Fig. \ref{pdf} shows the pdf considering a total time period from 2010 up to 2015, with $\tau\:=5,\:25,\:125,\:625$ and $3125$min., shifted vertically for clarity. All pdf exhibit a symmetric profile featured by long tails at large $\left|\delta p(\tau)\right|$, except at the largest $\tau$, where the pdf is Gaussian within the statistical noise. Similar tailed pdf have been observed in other price fluctuations in finance  \cite{Mantegna2007, Bouchaud2000a}. These pdf are characteristic to the particular system under study, the EURUSD in our case, and it does not change if different periods are studied; Fig. \ref{pdf}(a) compares the pdf for the years 2001 and 2007 (years with particular economic evolution), with the average in the period $2010-2015$.

Aiming to obtain a quantitative description of the price pdf, we borrow a model from glasses that has proven successful when describing data from experiments and simulations \cite{Kob2007}. Typically, tailed distributions in economy are approached by self-similarity or described through Levy flights combined with Gaussian ones \cite{Sanchez2015, Mantegna2007}, which are employed in few models, in particular the Cont-Bouchaud spin model \cite{Bouchaud2000b}, or the power law one \cite{Clauset09}. However, these require heuristic arguments or restrictions, such as a constant number of market investors or agents correlated according to trading strategies, proposed to be equal to spin domains. Our scope avoids considering these kind of conditions, as we directly study the EURUSD market as a whole, where any agent can behave freely.  In physical glasses, every particle is ideally caged by its own neighbours, restricting the structural relaxation of the whole system. Thermal fluctuations, however, allows particles to jump from one cage to another, on a large time scale. It is also useful to recall the classical description of glasses or undercooled fluids in terms of their free energy landscape containing multiple shallow minima (basins), separated by high barriers \cite{Goldstein69}. A similar model for financial markets, implying that the price is transiently trapped and eventually jumps out on large time scales, is attempted in the following, based on a simple model for particle glasses developed by Chaudri et al. \cite{Kob2007}.

In the model, particles are transiently trapped in cages, described by a time-independent Gaussian pdf ($f_{vib}(r)=(2\pi l^2)^{-3/2}exp(-r^2/2l^2)$ in their work). Long range jumps are possible, according to a Gaussian distribution $f_{jump}(r)=(2\pi d^2)^{-3/2}exp(-r^2/2d^2)$, on a large time scale. The probability to jump for the first time is given by an exponential distribution $\phi_{1}(t) = \tau_{1}^{-1}exp(-t/\tau_{1})$ while subsequent jumps occur faster according to $\phi_{2}(t) = \tau_{2}^{-1}exp(-t/\tau_{2})$, with $\tau_{2}<\tau_{1}$. The overall displacement distribution, or van Hove function, $G(r,t)$ depicts the probability of finding a particle in $r$, at  time $t$, and it is calculated in the Fourier-Laplace domain, $G(q,s)$. Here, we have considered this van Hove distribution to analyze the EURUSD pdf, taking into consideration the new dimensionality of the problem, the scalar price instead of the position vector. Back transforming to price-time domain, the distribution reads:
\begin{gather}
  G(p, t) = \tau_{1}f_{vib}(p)\phi_{1}(t)+FT^{-1}\biggl[\tilde{f}_{vib}(q)\tilde{f}(q)\tau_{2}\times\nonumber\\
  \times\frac{\exp\lbrace(\tilde{f}(q)-1)t/\tau_{2}\rbrace-\exp(-t/\tau_{1})}{\tau_{2}-\tau_{1}+\tilde{f}(q)\tau_{1}}\biggr] \label{model1}
\end{gather}
Here $\tilde{f}(q) = \tilde{f}_{vib}(q)\tilde{f}_{jump}(q)$, $\tilde{f}(q)$ is the Fourier transform of function $f(p)$, $q$ is the conjugate variable of price $p$ in the Fourier space and $FT^{-1}$ denotes the Inverse Fourier Transform. In this original model, the particle is assumed to explore its cage on a time scale much shorter than $\tau_1$ or $\tau_2$, thus $f_{vib}(r)$ is time independent. Because this assumption can not be made a priori in the EURUSD system, the model is modified to introduce short time diffusion, implying thus a finite time to explore the cage. For this purpose, we consider an Ornstein-Uhlenbeck process to calculate $f_{vib}(p, t)=\sqrt{\alpha/2\pi D(1-e^{-2\alpha t)})}\exp\{-\alpha p^{2}/2D(1 - e^{-2\alpha t})\}$ , with $D$ the diffusion coefficient and $\alpha=D/l^{2}$ \cite{Uhlenbeck1930, Gillespie1996, Risken1989}. This depicts a particle describing Brownian motion with a linear central force pulling it towards its origin. 

Using this new model, the experimental pdf are fitted, as shown in Fig. \ref{pdf} (lower panel), using $D$, $l$, $d$, $\tau_1$ and $\tau_2$, as fitting parameters, identical for all values of $\tau$. The values of the parameters are: $D=2\cdot 10^{-8}\,\mbox{min.}^{-1}$, $l=30\cdot 10^{-4},d=15\cdot 10^{-4},\tau_1=400\,\mbox{min.},\tau_2=300\,\mbox{min.}$, resulting in very good agreement with the experimental data, except for the lowest $\tau$. The comparison of the complementary cumulative distribution functions (cdf) from the model and experimental data, which is focused on the behaviour at long distances, is also satisfactory, as shown in the supplemental material \cite{SuppMat}. Note that parameters indicate the EURUSD price caged within intervals of ca. $0.30$ cents, and jumps out of this range occur on a time scale of approximately five hours.

\begin{figure}
  \includegraphics[scale=0.35]{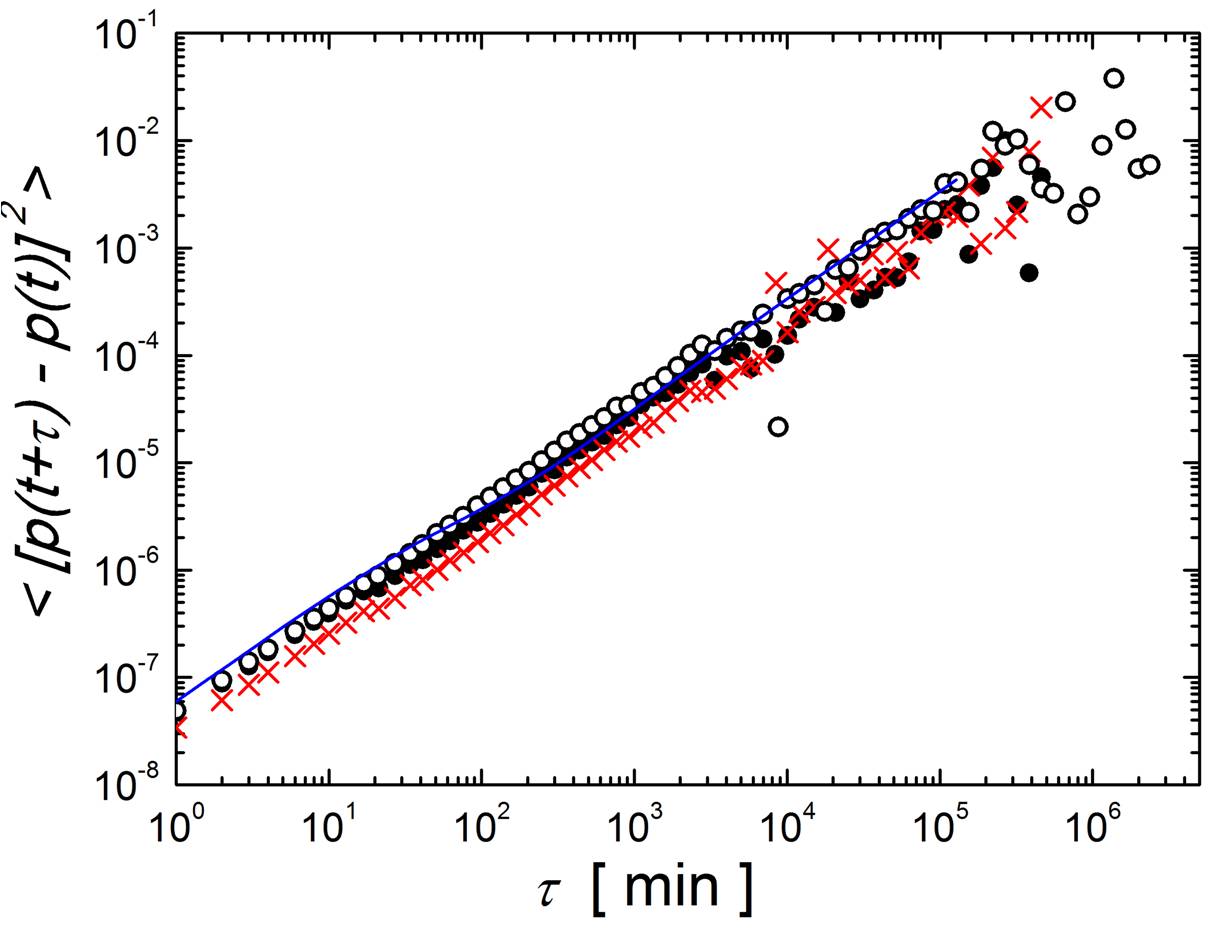}
  \caption{EURUSD MSPDs as calculated from 1min. experimental data for the period $2010-2015$ (open circles), to the year $2001$ (filled circles) and to the year $2007$ (crosses). The line is the MSPD according to the van Hove function of Eq. \ref{model1}, with the same parameters as Fig. \ref{pdf}b.\label{mspd}}
 \end{figure}

A hallmark of undercooled systems, as mentioned previously, is the separation between microscopic and structural dynamics, which results in correlation functions decaying in two steps, or intermediate plateaus in the tagged particle mean squared displacement (MSD)\cite{Lubelski2008, Scheffold2001, Weitz1993}. Analogously, we propose the mean squared price displacement (MSPD):
\begin{gather}
\big\langle\Delta p^{2}(\tau)\big\rangle = \big\langle[p(t_{0}+\tau) - p(t_{0})]^{2}\big\rangle
\label{mspd-eq}
\end{gather}
Despite fluctuation pdf being characteristic to arrested systems, the MSPD is linear, as indicated in Fig. \ref{mspd}. Even more, such diffusive behavior is found in all years, shown in Fig. \ref{mspd}, and all MSPDs exhibit a common origin, indicating that the diffusion coefficient, therefore market dynamics, is characteristic of the EURUSD regardless of the year considered, including years of conflict to economy, such as $2001$ or $2007$. In passing we note that short time dynamics is Brownian, what leads to establish the analogy with undercooled colloids, either glasses or gels, rather than atomic glasses. The MSPD calculated from our model is also presented in Fig. \ref{mspd}, correctly reproducing the data. The precise combination of the parameters $\{l,d,\tau_1,\tau_2\}$ and $D$ produces this linear evolution of $\Delta p^2(\tau)$ although the pdf differ clearly from Gaussian. 

\begin{figure}
  \includegraphics[scale=0.65]{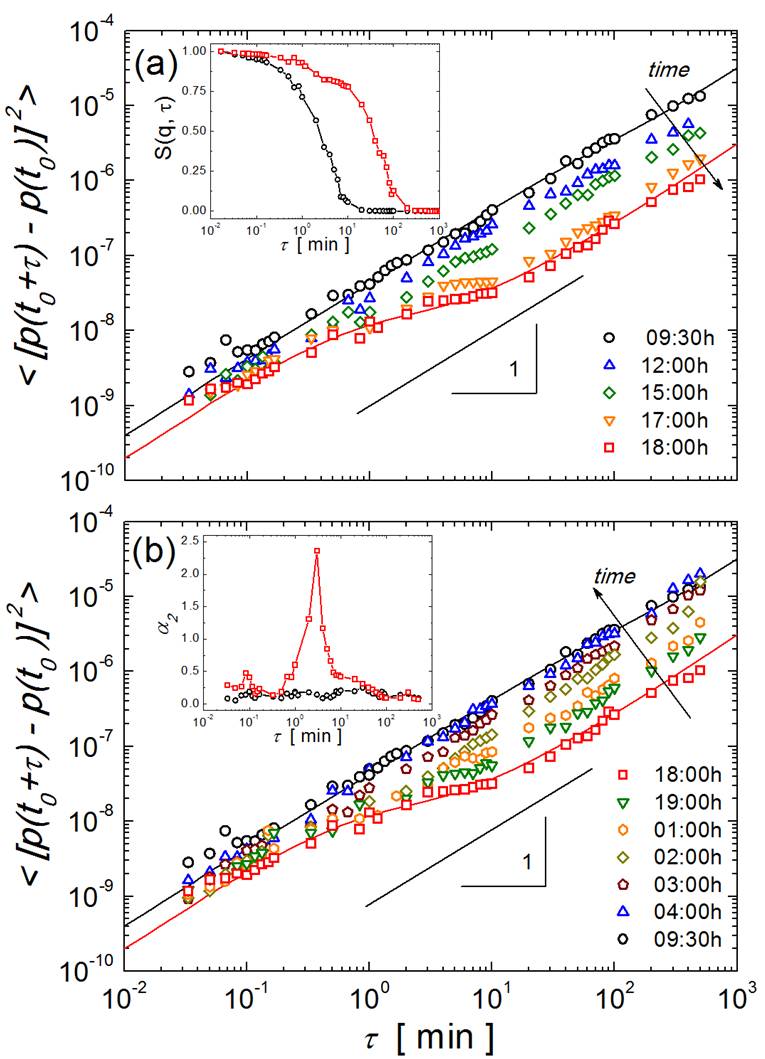}
  \caption{EURUSD MSPD as calculated from $2015$ experimental data by considering time periods of 24h (ET). In  the time range 9:30am - 6:00pm (a), a transition from diffusive to an undercooled dynamics is observed. Commencing times are labelled. By contrast, in the time frame of 6:00pm - 9:30am, the transition is reversed (b). The lines are the fitting of the model for the different cases. The insets show the price correlation function, $S(q, \tau)$, (a), and non-Gaussian parameter $\alpha_{2}$, (b).
  \label{msdp-24h}}
\end{figure}

Given this unexpected result, we seek a particular case or regime where the non-Gaussian components are important enough to produce deviations in the MSPD with respect to Brownian diffusion. Thus, we de-aggregate the data according to the commencing time, and study the dynamics for the next 24 hours. Additionally, we change the database to second-resolved prices, to access the short time dynamics. In particular, the year $2015$ is studied and we first set the starting point for the calculation of $\langle \delta p^2(\tau) \rangle$ at the opening of the New York Stock Exchange (NYSE), 9:30 am Eastern Time (ET). The MSPD for the forthcoming 24 hours after this commencing time, $t_0$, is averaged over all possible days. The same procedure is repeated for different values of $t_0$, covering the range of 24 hours. Recall that foreign exchange markets remain open continuously besides weekends. Results are presented in Fig. \ref{msdp-24h}, where a striking dependence of the MSPD with $t_0$ can be identified.

When the opening of the NYSE is taken as starting time, the dynamics is diffusive, but for later commencing times, the magnitude of the MSPD decreases and a shoulder appears at intermediate times, when $t_0$ approaches the closure of  NYSE. This trend continues developing and at $t_0 =$ 6:00pm ET the MSPD becomes minimal, exhibiting an initial diffusion-like increase at low $\tau$, crossing over to a quasi-plateau, and recovering again the linear behaviour for large $\tau$. This behaviour is close to the profile commonly found in arrested colloidal systems, either by aggregation, or at high particle density \cite{Wyss2001, Weeks2007}, due to the transient trapping of particles inside cages of neighbours (in glasses), or in a network of bonds (in gels). The same behaviour in the MSPD suggests that price dynamics considered at daily periods with the starting time between 3:00pm and 6:00pm ET undergoes to a temporary dynamic arrest. The transient trapping has a typical time scale of about $\tau = 30$min., and is followed by diffusive dynamics. Commencing times later than 6:00pm reverse this trend; MSPDs increase and the intermediate plateau vanishes, approaching diffusion when the starting point of the MSPD calculation is between 2:00am and 4:00am, with the maximum MSPD at 9:30am. This observation is consistent with the correlation of market behaviour with activity, as done by Ito and Hashimoto for the US Dollar - Japanese Yen market \cite{Ito2006}.

The theoretical model can rationalize this behaviour, as shown in Fig. 4, for the two extreme cases, $t_0=$9:30am and 6:00pm. To fit the model, $D$ and the time scales $\tau_1$ and $\tau_2$ where fixed for all $t_0$, as these are expected to be intrinsic to the EURUSD system, and only $l$ and $d$ are varied. The fits shown in the figure are obtained for $l=3\cdot 10^{-3}$ and $d=1.5\cdot 10^{-3}$ for $t_0=$ 9:30am, and $l=0.1\cdot 10^{-3}$ and $d=0.15\cdot 10^{-3}$ for $t_0=$ 6:00pm. Note that in the latter, both $l$ and $d$ are much smaller, but also $l<d$, characteristic of arrested systems, and the cage size is of order $\sim l$.

Given this similarity at the level of the MSPD, we study other properties which serve to identify undercooled states or glasses, in particular, the dynamic structure factor, $S(q, \tau) = \left<\exp\{iq(p(t_{0}+\tau)-p(t_{0}))\}\right>$, and the one-dimensional non-Gaussian parameter, $\alpha_{2} (\tau) = \left< \Delta p(\tau)^{4} \right>/3\left< \Delta p(\tau)^{2} \right>^{2} -1$. $S(q, \tau)$ starts from $1$ decaying to zero in fluids and to a finite value in glasses. In undercooled systems, it shows an intermediate plateau, that depends on the wavevector, $q$. For the EURUSD system, we select a value of $q\sim 2\pi/l$ to probe this range of price variations, and the resulting correlation functions are shown in the inset to Fig. \ref{msdp-24h}a, and further detailed at the supplemental material \cite{SuppMat}. The non-Gaussian parameter, $\alpha_2$, on the other hand, quantifies the deviation of the pdf from Gaussian. In undercooled fluids, $\alpha_2$ starts from zero (the pdf is Gaussian at short times), describes a maximum when the particles are caged and start to break free, to become zero again at long times. The non-Gaussian parameter for the EURUSD system, shown in the inset to Fig. \ref{msdp-24h}b and detailed at the supplemental material \cite{SuppMat}, indeed shows a maximum when $t_0=$6:00pm in the time range where the shoulder of the MSPD appears. Both $S(q,t)$ and $\alpha_2$ confirm the analogy between the dynamics of supercooled fluids and the EURUSD market.

The dynamics of economic markets is often described as diffusive, based on the linear MSPD profile when considering long periods, where the efficient market hypothesis (EMH) or fractal market hypothesis (FMH) can be invoked to explain such behavior \cite{Malkiel1973, Peters1989}. The EMH states that market prices are due to all market information being available, making it impossible to beat the market; such statement is in fact compatible with a random walk. We show in this study that this diffusion-like behaviour is found only when averages over long times are taken in the MSPD, $\langle\Delta p^{2}(\tau)\rangle$, but disappear when local time averages at $24$ hours periods are performed and where either diffusive or arrested dynamics are featured. The FMH states, on the other hand, that investment strategies converge when considering short time frames, arresting the market dynamics and making it more inefficient. Our analysis suggests a fluid-to-glass transition upon the choice of the reference MSPD time, implying that markets become either efficient or fractal-like, depending on their activity.

In particle systems, the different dynamical regimes typical of fluid-to-glass transitions are obtained switching a physical parameter that controls the interaction between particles. It must be then further sought, which is the origin of the non-trivial dynamics shown above and if the analogy can be exploited to identify the equivalent set of parameters that control market dynamics, such as traded volume, number of investors in the market or even policy decisions from regulatory institutions. Furthermore, the symmetry between colloidal systems and the EURUSD exchange market is as well found in other currency pairs, such as the EURCHF. A new approach appears, where exchange rate currency pairs can be regarded as colloidal systems and due to their coupled dynamics, the whole foreign exchange market can be considered a single undercooled system. This view and its natural extension to stock and other markets lays promising when modeling financial markets and needs to be further addressed.

\begin{acknowledgments}
We acknowledge \texttt{histdata.com} and \texttt{oanda.com} for providing all currency exchange data. Funding from UOC, project N11-6139473, aimed at enhancing submission to H2020 calls, (J. C.-R.), the Spanish Ministerio de Econom\'ia, Industria y Competitividad and the European Regional Development Fund (ERDF) under project No. FIS2015-69022-P and MTM2015-64373-P, are gratefully acknowledged.
\end{acknowledgments}  

%


\begin{thebibliography}{58}%
\makeatletter
\providecommand \@ifxundefined [1]{%
 \@ifx{#1\undefined}
}%
\providecommand \@ifnum [1]{%
 \ifnum #1\expandafter \@firstoftwo
 \else \expandafter \@secondoftwo
 \fi
}%
\providecommand \@ifx [1]{%
 \ifx #1\expandafter \@firstoftwo
 \else \expandafter \@secondoftwo
 \fi
}%
\providecommand \natexlab [1]{#1}%
\providecommand \enquote  [1]{``#1''}%
\providecommand \bibnamefont  [1]{#1}%
\providecommand \bibfnamefont [1]{#1}%
\providecommand \citenamefont [1]{#1}%
\providecommand \href@noop [0]{\@secondoftwo}%
\providecommand \href [0]{\begingroup \@sanitize@url \@href}%
\providecommand \@href[1]{\@@startlink{#1}\@@href}%
\providecommand \@@href[1]{\endgroup#1\@@endlink}%
\providecommand \@sanitize@url [0]{\catcode `\\12\catcode `\$12\catcode
  `\&12\catcode `\#12\catcode `\^12\catcode `\_12\catcode `\%12\relax}%
\providecommand \@@startlink[1]{}%
\providecommand \@@endlink[0]{}%
\providecommand \url  [0]{\begingroup\@sanitize@url \@url }%
\providecommand \@url [1]{\endgroup\@href {#1}{\urlprefix }}%
\providecommand \urlprefix  [0]{URL }%
\providecommand \Eprint [0]{\href }%
\providecommand \doibase [0]{http://dx.doi.org/}%
\providecommand \selectlanguage [0]{\@gobble}%
\providecommand \bibinfo  [0]{\@secondoftwo}%
\providecommand \bibfield  [0]{\@secondoftwo}%
\providecommand \translation [1]{[#1]}%
\providecommand \BibitemOpen [0]{}%
\providecommand \bibitemStop [0]{}%
\providecommand \bibitemNoStop [0]{.\EOS\space}%
\providecommand \EOS [0]{\spacefactor3000\relax}%
\providecommand \BibitemShut  [1]{\csname bibitem#1\endcsname}%
\let\auto@bib@innerbib\@empty
\bibitem [{\citenamefont {Dove}(2003)}]{Dove2003}%
  \BibitemOpen
  \bibfield  {author} {\bibinfo {author} {\bibfnamefont {M.~T.}\ \bibnamefont
  {Dove}},\ }\href@noop {} {\emph {\bibinfo {title} {Structure and Dynamics: An
  Atomic View of Materials}}}\ (\bibinfo  {publisher} {Oxford University
  Press},\ \bibinfo {year} {2003})\BibitemShut {NoStop}%
\bibitem [{\citenamefont {Arimondo}\ \emph {et~al.}(2015)\citenamefont
  {Arimondo}, \citenamefont {Lin},\ and\ \citenamefont {Yelin}}]{Arimondo2015}%
  \BibitemOpen
  \bibinfo {editor} {\bibfnamefont {E.}~\bibnamefont {Arimondo}}, \bibinfo
  {editor} {\bibfnamefont {C.~C.}\ \bibnamefont {Lin}}, \ and\ \bibinfo
  {editor} {\bibfnamefont {S.~F.}\ \bibnamefont {Yelin}},\ eds.,\ \href@noop {}
  {\emph {\bibinfo {title} {Advances in Atomic, Molecular, and Optical
  Physics}}}\ (\bibinfo  {publisher} {Academic Press Inc},\ \bibinfo {year}
  {2015})\BibitemShut {NoStop}%
\bibitem [{\citenamefont {Giordano}\ and\ \citenamefont
  {Ruta}(2016)}]{Giordano2016}%
  \BibitemOpen
  \bibfield  {author} {\bibinfo {author} {\bibfnamefont {V.~M.}\ \bibnamefont
  {Giordano}}\ and\ \bibinfo {author} {\bibfnamefont {B.}~\bibnamefont
  {Ruta}},\ }\href@noop {} {\bibfield  {journal} {\bibinfo  {journal} {Nat.
  Commun.}\ }\textbf {\bibinfo {volume} {7}},\ \bibinfo {pages} {10344}
  (\bibinfo {year} {2016})}\BibitemShut {NoStop}%
\bibitem [{\citenamefont {Franklin}\ and\ \citenamefont
  {Shattuck}(2015)}]{Franklin2015}%
  \BibitemOpen
  \bibinfo {editor} {\bibfnamefont {S.~V.}\ \bibnamefont {Franklin}}\ and\
  \bibinfo {editor} {\bibfnamefont {M.~D.}\ \bibnamefont {Shattuck}},\ eds.,\
  \href@noop {} {\emph {\bibinfo {title} {Handbook of Granular Materials}}}\
  (\bibinfo  {publisher} {CRC Press},\ \bibinfo {year} {2015})\BibitemShut
  {NoStop}%
\bibitem [{\citenamefont {Josserand}\ \emph {et~al.}(2000)\citenamefont
  {Josserand}, \citenamefont {Tkachenko}, \citenamefont {Mueth},\ and\
  \citenamefont {Jaeger}}]{Josserand2000}%
  \BibitemOpen
  \bibfield  {author} {\bibinfo {author} {\bibfnamefont {C.}~\bibnamefont
  {Josserand}}, \bibinfo {author} {\bibfnamefont {A.~V.}\ \bibnamefont
  {Tkachenko}}, \bibinfo {author} {\bibfnamefont {D.~M.}\ \bibnamefont
  {Mueth}}, \ and\ \bibinfo {author} {\bibfnamefont {H.~M.}\ \bibnamefont
  {Jaeger}},\ }\href@noop {} {\bibfield  {journal} {\bibinfo  {journal} {Phys.
  Rev. Lett.}\ }\textbf {\bibinfo {volume} {85}},\ \bibinfo {pages} {3632}
  (\bibinfo {year} {2000})}\BibitemShut {NoStop}%
\bibitem [{\citenamefont {Liu}\ and\ \citenamefont {Nagel}(2001)}]{Liu2003}%
  \BibitemOpen
  \bibinfo {editor} {\bibfnamefont {A.~J.}\ \bibnamefont {Liu}}\ and\ \bibinfo
  {editor} {\bibfnamefont {S.~R.}\ \bibnamefont {Nagel}},\ eds.,\ \href@noop {}
  {\emph {\bibinfo {title} {Jamming and Rheology}}}\ (\bibinfo  {publisher}
  {CRC Press},\ \bibinfo {year} {2001})\BibitemShut {NoStop}%
\bibitem [{\citenamefont {Cipelletti}\ and\ \citenamefont
  {Ramos}(2002)}]{Cipelletti2002}%
  \BibitemOpen
  \bibfield  {author} {\bibinfo {author} {\bibfnamefont {L.}~\bibnamefont
  {Cipelletti}}\ and\ \bibinfo {author} {\bibfnamefont {L.}~\bibnamefont
  {Ramos}},\ }\href@noop {} {\bibfield  {journal} {\bibinfo  {journal} {Curr.
  Op. Coll. Interf. Sci.}\ }\textbf {\bibinfo {volume} {7}},\ \bibinfo {pages}
  {228} (\bibinfo {year} {2002})}\BibitemShut {NoStop}%
\bibitem [{\citenamefont {de~Vegvar}\ and\ \citenamefont
  {Fulton}(1993)}]{deVegvar1993}%
  \BibitemOpen
  \bibfield  {author} {\bibinfo {author} {\bibfnamefont {P.~G.~N.}\
  \bibnamefont {de~Vegvar}}\ and\ \bibinfo {author} {\bibfnamefont {T.~A.}\
  \bibnamefont {Fulton}},\ }\href@noop {} {\bibfield  {journal} {\bibinfo
  {journal} {Phys. Rev. Lett.}\ }\textbf {\bibinfo {volume} {71}},\ \bibinfo
  {pages} {3537} (\bibinfo {year} {1993})}\BibitemShut {NoStop}%
\bibitem [{\citenamefont {Castillo}\ and\ \citenamefont
  {Parsaeian}(2007)}]{Castillo2007}%
  \BibitemOpen
  \bibfield  {author} {\bibinfo {author} {\bibfnamefont {H.~E.}\ \bibnamefont
  {Castillo}}\ and\ \bibinfo {author} {\bibfnamefont {A.}~\bibnamefont
  {Parsaeian}},\ }\href@noop {} {\bibfield  {journal} {\bibinfo  {journal}
  {Nat. Phys.}\ }\textbf {\bibinfo {volume} {3}},\ \bibinfo {pages} {26}
  (\bibinfo {year} {2007})}\BibitemShut {NoStop}%
\bibitem [{\citenamefont {Rygbyt}\ and\ \citenamefont
  {Roe}(1990)}]{Rigbyt1990}%
  \BibitemOpen
  \bibfield  {author} {\bibinfo {author} {\bibfnamefont {D.}~\bibnamefont
  {Rygbyt}}\ and\ \bibinfo {author} {\bibfnamefont {R.~J.}\ \bibnamefont
  {Roe}},\ }\href@noop {} {\bibfield  {journal} {\bibinfo  {journal}
  {Macromol.}\ }\textbf {\bibinfo {volume} {23}},\ \bibinfo {pages} {5312}
  (\bibinfo {year} {1990})}\BibitemShut {NoStop}%
\bibitem [{\citenamefont {Gardel}\ \emph {et~al.}(2009)\citenamefont {Gardel},
  \citenamefont {Sitaridou}, \citenamefont {Facto}, \citenamefont {Keene},
  \citenamefont {Hattam}, \citenamefont {Easwar},\ and\ \citenamefont
  {Menon}}]{Gardel2009}%
  \BibitemOpen
  \bibfield  {author} {\bibinfo {author} {\bibfnamefont {E.}~\bibnamefont
  {Gardel}}, \bibinfo {author} {\bibfnamefont {E.}~\bibnamefont {Sitaridou}},
  \bibinfo {author} {\bibfnamefont {K.}~\bibnamefont {Facto}}, \bibinfo
  {author} {\bibfnamefont {E.}~\bibnamefont {Keene}}, \bibinfo {author}
  {\bibfnamefont {K.}~\bibnamefont {Hattam}}, \bibinfo {author} {\bibfnamefont
  {N.}~\bibnamefont {Easwar}}, \ and\ \bibinfo {author} {\bibfnamefont
  {N.}~\bibnamefont {Menon}},\ }\href@noop {} {\bibfield  {journal} {\bibinfo
  {journal} {Phil. Trans. R. Soc. A}\ }\textbf {\bibinfo {volume} {367}},\
  \bibinfo {pages} {5109} (\bibinfo {year} {2009})}\BibitemShut {NoStop}%
\bibitem [{\citenamefont {Duri}\ \emph {et~al.}(2005)\citenamefont {Duri},
  \citenamefont {Bissig}, \citenamefont {Trappe},\ and\ \citenamefont
  {Cipelletti}}]{Duri2005}%
  \BibitemOpen
  \bibfield  {author} {\bibinfo {author} {\bibfnamefont {A.}~\bibnamefont
  {Duri}}, \bibinfo {author} {\bibfnamefont {H.}~\bibnamefont {Bissig}},
  \bibinfo {author} {\bibfnamefont {V.}~\bibnamefont {Trappe}}, \ and\ \bibinfo
  {author} {\bibfnamefont {L.}~\bibnamefont {Cipelletti}},\ }\href@noop {}
  {\bibfield  {journal} {\bibinfo  {journal} {Phys. Rev. E}\ }\textbf {\bibinfo
  {volume} {72}},\ \bibinfo {pages} {051401} (\bibinfo {year}
  {2005})}\BibitemShut {NoStop}%
\bibitem [{\citenamefont {Sessoms}\ \emph {et~al.}(2010)\citenamefont
  {Sessoms}, \citenamefont {Bissig}, \citenamefont {Duri}, \citenamefont
  {Cipelletti},\ and\ \citenamefont {Trappe}}]{Sessoms2010}%
  \BibitemOpen
  \bibfield  {author} {\bibinfo {author} {\bibfnamefont {D.~A.}\ \bibnamefont
  {Sessoms}}, \bibinfo {author} {\bibfnamefont {H.}~\bibnamefont {Bissig}},
  \bibinfo {author} {\bibfnamefont {A.}~\bibnamefont {Duri}}, \bibinfo {author}
  {\bibfnamefont {L.}~\bibnamefont {Cipelletti}}, \ and\ \bibinfo {author}
  {\bibfnamefont {V.}~\bibnamefont {Trappe}},\ }\href@noop {} {\bibfield
  {journal} {\bibinfo  {journal} {Royal Soc. of Chem.}\ }\textbf {\bibinfo
  {volume} {6}},\ \bibinfo {pages} {3030} (\bibinfo {year} {2010})}\BibitemShut
  {NoStop}%
\bibitem [{\citenamefont {Chaudhuri}\ \emph {et~al.}(2007)\citenamefont
  {Chaudhuri}, \citenamefont {Berthier},\ and\ \citenamefont {Kob}}]{Kob2007}%
  \BibitemOpen
  \bibfield  {author} {\bibinfo {author} {\bibfnamefont {P.}~\bibnamefont
  {Chaudhuri}}, \bibinfo {author} {\bibfnamefont {L.}~\bibnamefont {Berthier}},
  \ and\ \bibinfo {author} {\bibfnamefont {W.}~\bibnamefont {Kob}},\
  }\href@noop {} {\bibfield  {journal} {\bibinfo  {journal} {Phys. Rev. Lett.}\
  }\textbf {\bibinfo {volume} {99}},\ \bibinfo {pages} {060604} (\bibinfo
  {year} {2007})}\BibitemShut {NoStop}%
\bibitem [{\citenamefont {March}\ and\ \citenamefont {Tosi}(1991)}]{March1991}%
  \BibitemOpen
  \bibfield  {author} {\bibinfo {author} {\bibfnamefont {N.~H.}\ \bibnamefont
  {March}}\ and\ \bibinfo {author} {\bibfnamefont {M.~P.}\ \bibnamefont
  {Tosi}},\ }\href@noop {} {\emph {\bibinfo {title} {Atomic Dynamics in
  Liquids}}}\ (\bibinfo  {publisher} {Dover Pub. Inc.},\ \bibinfo {year}
  {1991})\BibitemShut {NoStop}%
\bibitem [{\citenamefont {Fernandez-Nieves}\ and\ \citenamefont
  {Puertas}(2016)}]{AFN-AMP2016}%
  \BibitemOpen
  \bibinfo {editor} {\bibfnamefont {A.}~\bibnamefont {Fernandez-Nieves}}\ and\
  \bibinfo {editor} {\bibfnamefont {A.~M.}\ \bibnamefont {Puertas}},\ eds.,\
  \href@noop {} {\emph {\bibinfo {title} {Fluids, Colloids and Soft Materials:
  An introduction to Soft Matter Physics}}}\ (\bibinfo  {publisher} {Wiley},\
  \bibinfo {year} {2016})\BibitemShut {NoStop}%
\bibitem [{\citenamefont {Jones}(2002)}]{Jones2002}%
  \BibitemOpen
  \bibfield  {author} {\bibinfo {author} {\bibfnamefont {R.~A.~L.}\
  \bibnamefont {Jones}},\ }\href@noop {} {\emph {\bibinfo {title} {Soft
  Condensed Matter}}}\ (\bibinfo  {publisher} {Oxford University Press},\
  \bibinfo {year} {2002})\BibitemShut {NoStop}%
\bibitem [{\citenamefont {Pusey}\ and\ \citenamefont {van
  Megen}(1986)}]{Pusey1986}%
  \BibitemOpen
  \bibfield  {author} {\bibinfo {author} {\bibfnamefont {P.~N.}\ \bibnamefont
  {Pusey}}\ and\ \bibinfo {author} {\bibfnamefont {W.}~\bibnamefont {van
  Megen}},\ }\href@noop {} {\bibfield  {journal} {\bibinfo  {journal} {Nature}\
  }\textbf {\bibinfo {volume} {320}},\ \bibinfo {pages} {340} (\bibinfo {year}
  {1986})}\BibitemShut {NoStop}%
\bibitem [{\citenamefont {Crocker}\ and\ \citenamefont
  {Grier}(1994)}]{Crocker1994}%
  \BibitemOpen
  \bibfield  {author} {\bibinfo {author} {\bibfnamefont {J.~C.}\ \bibnamefont
  {Crocker}}\ and\ \bibinfo {author} {\bibfnamefont {D.~G.}\ \bibnamefont
  {Grier}},\ }\href@noop {} {\bibfield  {journal} {\bibinfo  {journal} {Phys.
  Rev. Lett.}\ }\textbf {\bibinfo {volume} {73}},\ \bibinfo {pages} {352}
  (\bibinfo {year} {1994})}\BibitemShut {NoStop}%
\bibitem [{\citenamefont {Pham}\ \emph {et~al.}(2002)\citenamefont {Pham},
  \citenamefont {Puertas}, \citenamefont {Berenholtz}, \citenamefont
  {Egelhaaf}, \citenamefont {Moussaïd}, \citenamefont {Pusey}, \citenamefont
  {Schofield}, \citenamefont {Cates}, \citenamefont {Fuchs},\ and\
  \citenamefont {Poon}}]{Pham2002}%
  \BibitemOpen
  \bibfield  {author} {\bibinfo {author} {\bibfnamefont {K.~N.}\ \bibnamefont
  {Pham}}, \bibinfo {author} {\bibfnamefont {A.~M.}\ \bibnamefont {Puertas}},
  \bibinfo {author} {\bibfnamefont {J.}~\bibnamefont {Berenholtz}}, \bibinfo
  {author} {\bibfnamefont {S.~U.}\ \bibnamefont {Egelhaaf}}, \bibinfo {author}
  {\bibfnamefont {A.}~\bibnamefont {Moussaïd}}, \bibinfo {author}
  {\bibfnamefont {P.~N.}\ \bibnamefont {Pusey}}, \bibinfo {author}
  {\bibfnamefont {A.~B.}\ \bibnamefont {Schofield}}, \bibinfo {author}
  {\bibfnamefont {M.~E.}\ \bibnamefont {Cates}}, \bibinfo {author}
  {\bibfnamefont {M.}~\bibnamefont {Fuchs}}, \ and\ \bibinfo {author}
  {\bibfnamefont {W.~C.~K.}\ \bibnamefont {Poon}},\ }\href@noop {} {\bibfield
  {journal} {\bibinfo  {journal} {Science}\ }\textbf {\bibinfo {volume}
  {296}},\ \bibinfo {pages} {104} (\bibinfo {year} {2002})}\BibitemShut
  {NoStop}%
\bibitem [{\citenamefont {Poon}(2004)}]{Poon2004}%
  \BibitemOpen
  \bibfield  {author} {\bibinfo {author} {\bibfnamefont {W.}~\bibnamefont
  {Poon}},\ }\href@noop {} {\bibfield  {journal} {\bibinfo  {journal}
  {Science}\ }\textbf {\bibinfo {volume} {304}},\ \bibinfo {pages} {830}
  (\bibinfo {year} {2004})}\BibitemShut {NoStop}%
\bibitem [{\citenamefont {Meyers}(2011)}]{Meyers2011}%
  \BibitemOpen
  \bibinfo {editor} {\bibfnamefont {R.~A.}\ \bibnamefont {Meyers}},\ ed.,\
  \href@noop {} {\emph {\bibinfo {title} {Complex Systems in Finance and
  Econometrics}}}\ (\bibinfo  {publisher} {Springer},\ \bibinfo {year}
  {2011})\BibitemShut {NoStop}%
\bibitem [{\citenamefont {Bachelier}(2006)}]{Bachelier2006}%
  \BibitemOpen
  \bibfield  {author} {\bibinfo {author} {\bibfnamefont {L.}~\bibnamefont
  {Bachelier}},\ }\href@noop {} {\emph {\bibinfo {title} {Theory of
  Speculation}}}\ (\bibinfo  {publisher} {Princeton University Press},\
  \bibinfo {year} {2006})\BibitemShut {NoStop}%
\bibitem [{\citenamefont {Fama}\ and\ \citenamefont {Miller}(1972)}]{Fama1972}%
  \BibitemOpen
  \bibfield  {author} {\bibinfo {author} {\bibfnamefont {E.~F.}\ \bibnamefont
  {Fama}}\ and\ \bibinfo {author} {\bibfnamefont {M.~H.}\ \bibnamefont
  {Miller}},\ }\href@noop {} {\emph {\bibinfo {title} {The Theory of
  Finance}}}\ (\bibinfo  {publisher} {Holt Rinehart and Winston},\ \bibinfo
  {year} {1972})\BibitemShut {NoStop}%
\bibitem [{\citenamefont {Fama}(1976)}]{Fama1976}%
  \BibitemOpen
  \bibfield  {author} {\bibinfo {author} {\bibfnamefont {E.~F.}\ \bibnamefont
  {Fama}},\ }\href@noop {} {\emph {\bibinfo {title} {Foundations of Finance,
  Portfolio Decision and Securities Prices}}}\ (\bibinfo  {publisher} {Basic
  Books},\ \bibinfo {year} {1976})\BibitemShut {NoStop}%
\bibitem [{\citenamefont {Bouchaud}\ and\ \citenamefont
  {Potters}(2000)}]{Bouchaud2000a}%
  \BibitemOpen
  \bibfield  {author} {\bibinfo {author} {\bibfnamefont {J.-P.}\ \bibnamefont
  {Bouchaud}}\ and\ \bibinfo {author} {\bibfnamefont {M.}~\bibnamefont
  {Potters}},\ }\href@noop {} {\emph {\bibinfo {title} {Theory of Financial
  Risks}}}\ (\bibinfo  {publisher} {Cambridge Univ. Press},\ \bibinfo {year}
  {2000})\BibitemShut {NoStop}%
\bibitem [{\citenamefont {Mandelbrot}(1997)}]{Mandelbrot1997}%
  \BibitemOpen
  \bibfield  {author} {\bibinfo {author} {\bibfnamefont {B.~B.}\ \bibnamefont
  {Mandelbrot}},\ }\href@noop {} {\emph {\bibinfo {title} {Fractals and Scaling
  in Finance}}}\ (\bibinfo  {publisher} {Springer New York},\ \bibinfo {year}
  {1997})\BibitemShut {NoStop}%
\bibitem [{\citenamefont {Peters}(1996)}]{Peters1996}%
  \BibitemOpen
  \bibfield  {author} {\bibinfo {author} {\bibfnamefont {E.~E.}\ \bibnamefont
  {Peters}},\ }\href@noop {} {\emph {\bibinfo {title} {Fractal Market
  Analysis}}}\ (\bibinfo  {publisher} {Wiley},\ \bibinfo {year}
  {1996})\BibitemShut {NoStop}%
\bibitem [{\citenamefont {Stanley}\ \emph {et~al.}(1996)\citenamefont
  {Stanley}, \citenamefont {Afanasyev}, \citenamefont {Amaral}, \citenamefont
  {Buldyrev}, \citenamefont {Goldberger}, \citenamefont {Havlin}, \citenamefont
  {Leschhorn}, \citenamefont {Maass}, \citenamefont {Mantegna}, \citenamefont
  {Peng}, \citenamefont {Prince}, \citenamefont {Salinger}, \citenamefont
  {Stanley},\ and\ \citenamefont {Viswanathan}}]{Stanley96}%
  \BibitemOpen
  \bibfield  {author} {\bibinfo {author} {\bibfnamefont {H.}~\bibnamefont
  {Stanley}}, \bibinfo {author} {\bibfnamefont {V.}~\bibnamefont {Afanasyev}},
  \bibinfo {author} {\bibfnamefont {L.~A.~N.}\ \bibnamefont {Amaral}}, \bibinfo
  {author} {\bibfnamefont {S.}~\bibnamefont {Buldyrev}}, \bibinfo {author}
  {\bibfnamefont {A.}~\bibnamefont {Goldberger}}, \bibinfo {author}
  {\bibfnamefont {S.}~\bibnamefont {Havlin}}, \bibinfo {author} {\bibfnamefont
  {H.}~\bibnamefont {Leschhorn}}, \bibinfo {author} {\bibfnamefont
  {P.}~\bibnamefont {Maass}}, \bibinfo {author} {\bibfnamefont
  {R.}~\bibnamefont {Mantegna}}, \bibinfo {author} {\bibfnamefont {C.-K.}\
  \bibnamefont {Peng}}, \bibinfo {author} {\bibfnamefont {P.~A.}\ \bibnamefont
  {Prince}}, \bibinfo {author} {\bibfnamefont {M.~A.}\ \bibnamefont
  {Salinger}}, \bibinfo {author} {\bibfnamefont {M.}~\bibnamefont {Stanley}}, \
  and\ \bibinfo {author} {\bibfnamefont {G.}~\bibnamefont {Viswanathan}},\
  }\href@noop {} {\bibfield  {journal} {\bibinfo  {journal} {Physica A}\
  }\textbf {\bibinfo {volume} {224}},\ \bibinfo {pages} {302} (\bibinfo {year}
  {1996})}\BibitemShut {NoStop}%
\bibitem [{\citenamefont {Black}\ and\ \citenamefont {Scholes}(1973)}]{BS1973}%
  \BibitemOpen
  \bibfield  {author} {\bibinfo {author} {\bibfnamefont {F.}~\bibnamefont
  {Black}}\ and\ \bibinfo {author} {\bibfnamefont {M.}~\bibnamefont
  {Scholes}},\ }\href@noop {} {\bibfield  {journal} {\bibinfo  {journal} {J.
  Pol. Ec.}\ }\textbf {\bibinfo {volume} {81}},\ \bibinfo {pages} {637}
  (\bibinfo {year} {1973})}\BibitemShut {NoStop}%
\bibitem [{\citenamefont {Mantegna}\ and\ \citenamefont
  {Stanley}(1996)}]{Mantegna1996}%
  \BibitemOpen
  \bibfield  {author} {\bibinfo {author} {\bibfnamefont {R.~N.}\ \bibnamefont
  {Mantegna}}\ and\ \bibinfo {author} {\bibfnamefont {H.~E.}\ \bibnamefont
  {Stanley}},\ }\href@noop {} {\bibfield  {journal} {\bibinfo  {journal}
  {Nature}\ }\textbf {\bibinfo {volume} {383}},\ \bibinfo {pages} {587}
  (\bibinfo {year} {1996})}\BibitemShut {NoStop}%
\bibitem [{\citenamefont {Francq}\ and\ \citenamefont
  {Zakolan}(2010)}]{Francq2010}%
  \BibitemOpen
  \bibfield  {author} {\bibinfo {author} {\bibfnamefont {C.}~\bibnamefont
  {Francq}}\ and\ \bibinfo {author} {\bibfnamefont {J.-M.}\ \bibnamefont
  {Zakolan}},\ }\href@noop {} {\emph {\bibinfo {title} {GARCH Models:
  Structure, Statistical Inference and Financial Applications}}}\ (\bibinfo
  {publisher} {Wiley},\ \bibinfo {year} {2010})\BibitemShut {NoStop}%
\bibitem [{\citenamefont {Mantegna}\ and\ \citenamefont
  {Stanley}(2007)}]{Mantegna2007}%
  \BibitemOpen
  \bibfield  {author} {\bibinfo {author} {\bibfnamefont {R.~N.}\ \bibnamefont
  {Mantegna}}\ and\ \bibinfo {author} {\bibfnamefont {H.~E.}\ \bibnamefont
  {Stanley}},\ }\href@noop {} {\emph {\bibinfo {title} {Introduction to
  Econophysics: Correlations and Complexity in Finance}}}\ (\bibinfo
  {publisher} {Cambridge University Press},\ \bibinfo {year}
  {2007})\BibitemShut {NoStop}%
\bibitem [{\citenamefont {Richmond}\ \emph {et~al.}(2013)\citenamefont
  {Richmond}, \citenamefont {Mimkes},\ and\ \citenamefont
  {Hutzler}}]{Richmond2013}%
  \BibitemOpen
  \bibfield  {author} {\bibinfo {author} {\bibfnamefont {P.}~\bibnamefont
  {Richmond}}, \bibinfo {author} {\bibfnamefont {J.}~\bibnamefont {Mimkes}}, \
  and\ \bibinfo {author} {\bibfnamefont {S.}~\bibnamefont {Hutzler}},\
  }\href@noop {} {\emph {\bibinfo {title} {Econophysics and Physical
  Economics}}}\ (\bibinfo  {publisher} {Oxford University Press},\ \bibinfo
  {year} {2013})\BibitemShut {NoStop}%
\bibitem [{\citenamefont {Puertas}\ \emph {et~al.}(2002)\citenamefont
  {Puertas}, \citenamefont {Fuchs},\ and\ \citenamefont {Cates}}]{Puertas2002}%
  \BibitemOpen
  \bibfield  {author} {\bibinfo {author} {\bibfnamefont {A.~M.}\ \bibnamefont
  {Puertas}}, \bibinfo {author} {\bibfnamefont {M.}~\bibnamefont {Fuchs}}, \
  and\ \bibinfo {author} {\bibfnamefont {M.~E.}\ \bibnamefont {Cates}},\
  }\href@noop {} {\bibfield  {journal} {\bibinfo  {journal} {Phys. Rev. Lett.}\
  }\textbf {\bibinfo {volume} {88}},\ \bibinfo {pages} {098301} (\bibinfo
  {year} {2002})}\BibitemShut {NoStop}%
\bibitem [{\citenamefont {Puertas}\ \emph {et~al.}(2007)\citenamefont
  {Puertas}, \citenamefont {Fuchs},\ and\ \citenamefont {Cates}}]{Puertas2007}%
  \BibitemOpen
  \bibfield  {author} {\bibinfo {author} {\bibfnamefont {A.~M.}\ \bibnamefont
  {Puertas}}, \bibinfo {author} {\bibfnamefont {M.}~\bibnamefont {Fuchs}}, \
  and\ \bibinfo {author} {\bibfnamefont {M.~E.}\ \bibnamefont {Cates}},\
  }\href@noop {} {\bibfield  {journal} {\bibinfo  {journal} {Phys. Rev. E}\
  }\textbf {\bibinfo {volume} {75}},\ \bibinfo {pages} {031401} (\bibinfo
  {year} {2007})}\BibitemShut {NoStop}%
\bibitem [{\citenamefont {Clara-Rahola}\ \emph {et~al.}(2014)\citenamefont
  {Clara-Rahola}, \citenamefont {Contreras-Caceres}, \citenamefont
  {Sierra-Martin}, \citenamefont {Maldonado-Valdivia}, \citenamefont {Hund},
  \citenamefont {Fery}, \citenamefont {Hellweg},\ and\ \citenamefont
  {Fernandez-Barbero}}]{Clara2014}%
  \BibitemOpen
  \bibfield  {author} {\bibinfo {author} {\bibfnamefont {J.}~\bibnamefont
  {Clara-Rahola}}, \bibinfo {author} {\bibfnamefont {R.}~\bibnamefont
  {Contreras-Caceres}}, \bibinfo {author} {\bibfnamefont {B.}~\bibnamefont
  {Sierra-Martin}}, \bibinfo {author} {\bibfnamefont {A.}~\bibnamefont
  {Maldonado-Valdivia}}, \bibinfo {author} {\bibfnamefont {M.}~\bibnamefont
  {Hund}}, \bibinfo {author} {\bibfnamefont {A.}~\bibnamefont {Fery}}, \bibinfo
  {author} {\bibfnamefont {T.}~\bibnamefont {Hellweg}}, \ and\ \bibinfo
  {author} {\bibfnamefont {A.}~\bibnamefont {Fernandez-Barbero}},\ }\href@noop
  {} {\bibfield  {journal} {\bibinfo  {journal} {Coll. Surf. A}\ }\textbf
  {\bibinfo {volume} {463}},\ \bibinfo {pages} {18} (\bibinfo {year}
  {2014})}\BibitemShut {NoStop}%
\bibitem [{\citenamefont {Clara-Rahola}\ \emph {et~al.}(2015)\citenamefont
  {Clara-Rahola}, \citenamefont {Brzinski}, \citenamefont {Semwogerere},
  \citenamefont {Feitosa}, \citenamefont {Crocker}, \citenamefont {Sato},
  \citenamefont {Breedveld},\ and\ \citenamefont {Weeks}}]{Clara2015}%
  \BibitemOpen
  \bibfield  {author} {\bibinfo {author} {\bibfnamefont {J.}~\bibnamefont
  {Clara-Rahola}}, \bibinfo {author} {\bibfnamefont {T.~A.}\ \bibnamefont
  {Brzinski}}, \bibinfo {author} {\bibfnamefont {D.}~\bibnamefont
  {Semwogerere}}, \bibinfo {author} {\bibfnamefont {K.}~\bibnamefont
  {Feitosa}}, \bibinfo {author} {\bibfnamefont {J.~C.}\ \bibnamefont
  {Crocker}}, \bibinfo {author} {\bibfnamefont {J.}~\bibnamefont {Sato}},
  \bibinfo {author} {\bibfnamefont {V.}~\bibnamefont {Breedveld}}, \ and\
  \bibinfo {author} {\bibfnamefont {E.~R.}\ \bibnamefont {Weeks}},\ }\href@noop
  {} {\bibfield  {journal} {\bibinfo  {journal} {Phys. Rev. E}\ }\textbf
  {\bibinfo {volume} {91}},\ \bibinfo {pages} {010301} (\bibinfo {year}
  {2015})}\BibitemShut {NoStop}%
\bibitem [{\citenamefont {Kaizoji}\ \emph {et~al.}(2002)\citenamefont
  {Kaizoji}, \citenamefont {Bornholdt},\ and\ \citenamefont
  {Fujiwara}}]{Kaizoji2002}%
  \BibitemOpen
  \bibfield  {author} {\bibinfo {author} {\bibfnamefont {T.}~\bibnamefont
  {Kaizoji}}, \bibinfo {author} {\bibfnamefont {S.}~\bibnamefont {Bornholdt}},
  \ and\ \bibinfo {author} {\bibfnamefont {Y.}~\bibnamefont {Fujiwara}},\
  }\href@noop {} {\bibfield  {journal} {\bibinfo  {journal} {Physica A}\
  }\textbf {\bibinfo {volume} {316}},\ \bibinfo {pages} {441} (\bibinfo {year}
  {2002})}\BibitemShut {NoStop}%
\bibitem [{\citenamefont {Duarte-Queiros}(2016)}]{Duarte2016}%
  \BibitemOpen
  \bibfield  {author} {\bibinfo {author} {\bibfnamefont {S.~M.}\ \bibnamefont
  {Duarte-Queiros}},\ }\href@noop {} {\bibfield  {journal} {\bibinfo  {journal}
  {Chaos, Sol. Frac.}\ }\textbf {\bibinfo {volume} {88}},\ \bibinfo {pages}
  {24} (\bibinfo {year} {2016})}\BibitemShut {NoStop}%
\bibitem [{\citenamefont {Stosic}\ \emph {et~al.}(2016)\citenamefont {Stosic},
  \citenamefont {Stosic}, \citenamefont {Ludermir}, \citenamefont
  {de~Oliveira},\ and\ \citenamefont {Stosic}}]{Stosic2016}%
  \BibitemOpen
  \bibfield  {author} {\bibinfo {author} {\bibfnamefont {D.}~\bibnamefont
  {Stosic}}, \bibinfo {author} {\bibfnamefont {D.}~\bibnamefont {Stosic}},
  \bibinfo {author} {\bibfnamefont {T.}~\bibnamefont {Ludermir}}, \bibinfo
  {author} {\bibfnamefont {W.}~\bibnamefont {de~Oliveira}}, \ and\ \bibinfo
  {author} {\bibfnamefont {T.}~\bibnamefont {Stosic}},\ }\href@noop {}
  {\bibfield  {journal} {\bibinfo  {journal} {Physica A}\ }\textbf {\bibinfo
  {volume} {449}},\ \bibinfo {pages} {233} (\bibinfo {year}
  {2016})}\BibitemShut {NoStop}%
\bibitem [{Not()}]{Note1}%
  \BibitemOpen
  \href@noop {} {}\bibinfo {note} {Note that in stock markets, the log return
  is usually studied, $r(\tau)=\langle \ln p(t_0+\tau)/p(t_0) \rangle$, to
  account for exponential growth or decrease of $p(t)$. In our case, however,
  the variation of $p(t)$ is small, yielding $r(\tau)\approx \delta
  p(\tau)/p(t_0)$.}\BibitemShut {Stop}%
\bibitem [{\citenamefont {Sanchez}\ \emph {et~al.}(2015)\citenamefont
  {Sanchez}, \citenamefont {Trinidad}, \citenamefont {Garcia},\ and\
  \citenamefont {Fernandez}}]{Sanchez2015}%
  \BibitemOpen
  \bibfield  {author} {\bibinfo {author} {\bibfnamefont {M.~A.}\ \bibnamefont
  {Sanchez}}, \bibinfo {author} {\bibfnamefont {J.~E.}\ \bibnamefont
  {Trinidad}}, \bibinfo {author} {\bibfnamefont {J.}~\bibnamefont {Garcia}}, \
  and\ \bibinfo {author} {\bibfnamefont {M.}~\bibnamefont {Fernandez}},\
  }\href@noop {} {\bibfield  {journal} {\bibinfo  {journal} {Plos One}\
  }\textbf {\bibinfo {volume} {10}},\ \bibinfo {pages} {e0127824} (\bibinfo
  {year} {2015})}\BibitemShut {NoStop}%
\bibitem [{\citenamefont {Cont}\ and\ \citenamefont
  {Bouchaud}(2000)}]{Bouchaud2000b}%
  \BibitemOpen
  \bibfield  {author} {\bibinfo {author} {\bibfnamefont {R.}~\bibnamefont
  {Cont}}\ and\ \bibinfo {author} {\bibfnamefont {J.-P.}\ \bibnamefont
  {Bouchaud}},\ }\href@noop {} {\bibfield  {journal} {\bibinfo  {journal}
  {Macroecon. Dyn.}\ }\textbf {\bibinfo {volume} {4}},\ \bibinfo {pages} {170}
  (\bibinfo {year} {2000})}\BibitemShut {NoStop}%
\bibitem [{\citenamefont {Clauset}\ \emph {et~al.}(2009)\citenamefont
  {Clauset}, \citenamefont {Shalizi},\ and\ \citenamefont
  {Newman}}]{Clauset09}%
  \BibitemOpen
  \bibfield  {author} {\bibinfo {author} {\bibfnamefont {A.}~\bibnamefont
  {Clauset}}, \bibinfo {author} {\bibfnamefont {C.~R.}\ \bibnamefont
  {Shalizi}}, \ and\ \bibinfo {author} {\bibfnamefont {M.~E.}\ \bibnamefont
  {Newman}},\ }\href@noop {} {\bibfield  {journal} {\bibinfo  {journal} {SIAM
  Review}\ }\textbf {\bibinfo {volume} {51}},\ \bibinfo {pages} {661} (\bibinfo
  {year} {2009})}\BibitemShut {NoStop}%
\bibitem [{\citenamefont {Goldstein}(1969)}]{Goldstein69}%
  \BibitemOpen
  \bibfield  {author} {\bibinfo {author} {\bibfnamefont {M.}~\bibnamefont
  {Goldstein}},\ }\href@noop {} {\bibfield  {journal} {\bibinfo  {journal} {J.
  Chem. Phys.}\ ,\ \bibinfo {pages} {3728–3739}} (\bibinfo {year}
  {1969})}\BibitemShut {NoStop}%
\bibitem [{\citenamefont {Uhlenbeck}\ and\ \citenamefont
  {Ornstein}(1930)}]{Uhlenbeck1930}%
  \BibitemOpen
  \bibfield  {author} {\bibinfo {author} {\bibfnamefont {G.~E.}\ \bibnamefont
  {Uhlenbeck}}\ and\ \bibinfo {author} {\bibfnamefont {L.~S.}\ \bibnamefont
  {Ornstein}},\ }\href@noop {} {\bibfield  {journal} {\bibinfo  {journal}
  {Phys. Rev.}\ }\textbf {\bibinfo {volume} {36}},\ \bibinfo {pages} {823}
  (\bibinfo {year} {1930})}\BibitemShut {NoStop}%
\bibitem [{\citenamefont {Gillespie}(1996)}]{Gillespie1996}%
  \BibitemOpen
  \bibfield  {author} {\bibinfo {author} {\bibfnamefont {D.~T.}\ \bibnamefont
  {Gillespie}},\ }\href@noop {} {\bibfield  {journal} {\bibinfo  {journal}
  {Phys. Rev. E}\ }\textbf {\bibinfo {volume} {54}},\ \bibinfo {pages} {2084}
  (\bibinfo {year} {1996})}\BibitemShut {NoStop}%
\bibitem [{\citenamefont {Risken}(1989)}]{Risken1989}%
  \BibitemOpen
  \bibfield  {author} {\bibinfo {author} {\bibfnamefont {H.}~\bibnamefont
  {Risken}},\ }\href@noop {} {\emph {\bibinfo {title} {The Fokker–Planck
  Equation: Method of Solution and Applications}}}\ (\bibinfo  {publisher}
  {Springer-Verlag New York},\ \bibinfo {year} {1989})\BibitemShut {NoStop}%
\bibitem [{Sup()}]{SuppMat}%
  \BibitemOpen
  \href@noop {} {}\bibinfo {note} {See Supplemental Material at [URL will be
  inserted by publisher], which includes Refs. [51-56], for details concerning complementary cdfs, non
  gaussian parameter, correlation functions and the fit to MSPDs according to
  the theoretical model proposed in this work.}\BibitemShut {Stop}%
\bibitem [{\citenamefont {Adler}\ \emph {et~al.}(1998)\citenamefont {Adler},
  \citenamefont {Feldman},\ and\ \citenamefont {Taqqu}}]{Adler1998}%
  \BibitemOpen
  \bibinfo {editor} {\bibfnamefont {R.~J.}\ \bibnamefont {Adler}}, \bibinfo
  {editor} {\bibfnamefont {R.~E.}\ \bibnamefont {Feldman}}, \ and\ \bibinfo
  {editor} {\bibfnamefont {M.~S.}\ \bibnamefont {Taqqu}},\ eds.,\ \href@noop {}
  {\emph {\bibinfo {title} {A practical guide to heavy tails}}}\ (\bibinfo
  {publisher} {Birkhauser},\ \bibinfo {year} {1998})\BibitemShut {NoStop}%
\bibitem [{\citenamefont {Binder}\ and\ \citenamefont {Kob}(2011)}]{Binder05}%
  \BibitemOpen
  \bibfield  {author} {\bibinfo {author} {\bibfnamefont {K.}~\bibnamefont
  {Binder}}\ and\ \bibinfo {author} {\bibfnamefont {W.}~\bibnamefont {Kob}},\
  }\href@noop {} {\emph {\bibinfo {title} {{Glassy Materials and Disordered
  Solids: An Introduction to Their Statistical Mechanics (Revised Edition)}}}}\
  (\bibinfo  {publisher} {World Scientific},\ \bibinfo {year}
  {2011})\BibitemShut {NoStop}%
\bibitem [{\citenamefont {Cipelletti}\ \emph {et~al.}(2000)\citenamefont
  {Cipelletti}, \citenamefont {Manley}, \citenamefont {Ball},\ and\
  \citenamefont {Weitz}}]{Cipelletti2000}%
  \BibitemOpen
  \bibfield  {author} {\bibinfo {author} {\bibfnamefont {L.}~\bibnamefont
  {Cipelletti}}, \bibinfo {author} {\bibfnamefont {S.}~\bibnamefont {Manley}},
  \bibinfo {author} {\bibfnamefont {R.~C.}\ \bibnamefont {Ball}}, \ and\
  \bibinfo {author} {\bibfnamefont {D.~A.}\ \bibnamefont {Weitz}},\ }\href@noop
  {} {\bibfield  {journal} {\bibinfo  {journal} {Phys. Rev. Lett.}\ }\textbf
  {\bibinfo {volume} {84}},\ \bibinfo {pages} {2275} (\bibinfo {year}
  {2000})}\BibitemShut {NoStop}%
\bibitem [{\citenamefont {Rahman}(1964)}]{Rahman1964}%
  \BibitemOpen
  \bibfield  {author} {\bibinfo {author} {\bibfnamefont {A.}~\bibnamefont
  {Rahman}},\ }\href@noop {} {\bibfield  {journal} {\bibinfo  {journal} {Phys.
  Rev.}\ }\textbf {\bibinfo {volume} {136}},\ \bibinfo {pages} {A405} (\bibinfo
  {year} {1964})}\BibitemShut {NoStop}%
\bibitem [{\citenamefont {Kob}\ \emph {et~al.}(1997)\citenamefont {Kob},
  \citenamefont {Donati}, \citenamefont {Plimpton}, \citenamefont {Poole},\
  and\ \citenamefont {Glotzer}}]{Kob1997}%
  \BibitemOpen
  \bibfield  {author} {\bibinfo {author} {\bibfnamefont {W.}~\bibnamefont
  {Kob}}, \bibinfo {author} {\bibfnamefont {C.}~\bibnamefont {Donati}},
  \bibinfo {author} {\bibfnamefont {S.~J.}\ \bibnamefont {Plimpton}}, \bibinfo
  {author} {\bibfnamefont {P.~H.}\ \bibnamefont {Poole}}, \ and\ \bibinfo
  {author} {\bibfnamefont {S.~C.}\ \bibnamefont {Glotzer}},\ }\href@noop {}
  {\bibfield  {journal} {\bibinfo  {journal} {Phys. Rev. Lett.}\ }\textbf
  {\bibinfo {volume} {79}},\ \bibinfo {pages} {2827} (\bibinfo {year}
  {1997})}\BibitemShut {NoStop}%
\bibitem [{\citenamefont {Weeks}\ and\ \citenamefont
  {Weitz}(2002)}]{Weeks2002}%
  \BibitemOpen
  \bibfield  {author} {\bibinfo {author} {\bibfnamefont {E.~R.}\ \bibnamefont
  {Weeks}}\ and\ \bibinfo {author} {\bibfnamefont {D.~A.}\ \bibnamefont
  {Weitz}},\ }\href@noop {} {\bibfield  {journal} {\bibinfo  {journal} {Chem.
  Phys.}\ }\textbf {\bibinfo {volume} {284}},\ \bibinfo {pages} {361} (\bibinfo
  {year} {2002})}\BibitemShut {NoStop}%
\bibitem [{\citenamefont {Lubelski}\ \emph {et~al.}(2008)\citenamefont
  {Lubelski}, \citenamefont {Sokolov},\ and\ \citenamefont
  {Klafter}}]{Lubelski2008}%
  \BibitemOpen
  \bibfield  {author} {\bibinfo {author} {\bibfnamefont {A.}~\bibnamefont
  {Lubelski}}, \bibinfo {author} {\bibfnamefont {I.~M.}\ \bibnamefont
  {Sokolov}}, \ and\ \bibinfo {author} {\bibfnamefont {J.}~\bibnamefont
  {Klafter}},\ }\href@noop {} {\bibfield  {journal} {\bibinfo  {journal} {Phys.
  Rev. Lett.}\ }\textbf {\bibinfo {volume} {100}},\ \bibinfo {pages} {250602}
  (\bibinfo {year} {2008})}\BibitemShut {NoStop}%
\bibitem [{\citenamefont {Scheffold}\ \emph {et~al.}(2001)\citenamefont
  {Scheffold}, \citenamefont {Skipetrov}, \citenamefont {Romer},\ and\
  \citenamefont {Schurtenberger}}]{Scheffold2001}%
  \BibitemOpen
  \bibfield  {author} {\bibinfo {author} {\bibfnamefont {F.}~\bibnamefont
  {Scheffold}}, \bibinfo {author} {\bibfnamefont {S.~E.}\ \bibnamefont
  {Skipetrov}}, \bibinfo {author} {\bibfnamefont {S.}~\bibnamefont {Romer}}, \
  and\ \bibinfo {author} {\bibfnamefont {P.}~\bibnamefont {Schurtenberger}},\
  }\href@noop {} {\bibfield  {journal} {\bibinfo  {journal} {Phys. Rev. E}\
  }\textbf {\bibinfo {volume} {63}},\ \bibinfo {pages} {061404} (\bibinfo
  {year} {2001})}\BibitemShut {NoStop}%
\bibitem [{\citenamefont {Weitz}\ and\ \citenamefont {Pine}(1993)}]{Weitz1993}%
  \BibitemOpen
  \bibfield  {author} {\bibinfo {author} {\bibfnamefont {D.~A.}\ \bibnamefont
  {Weitz}}\ and\ \bibinfo {author} {\bibfnamefont {D.~J.}\ \bibnamefont
  {Pine}},\ }\href@noop {} {\emph {\bibinfo {title} {Dynamic Light
  Scattering}}}\ (\bibinfo  {publisher} {Oxford University Press},\ \bibinfo
  {year} {1993})\BibitemShut {NoStop}%
\bibitem [{\citenamefont {Wyss}\ \emph {et~al.}(2001)\citenamefont {Wyss},
  \citenamefont {Romer}, \citenamefont {Scheffold}, \citenamefont
  {Schurtenberger},\ and\ \citenamefont {Gauckler}}]{Wyss2001}%
  \BibitemOpen
  \bibfield  {author} {\bibinfo {author} {\bibfnamefont {H.~M.}\ \bibnamefont
  {Wyss}}, \bibinfo {author} {\bibfnamefont {S.}~\bibnamefont {Romer}},
  \bibinfo {author} {\bibfnamefont {F.}~\bibnamefont {Scheffold}}, \bibinfo
  {author} {\bibfnamefont {P.}~\bibnamefont {Schurtenberger}}, \ and\ \bibinfo
  {author} {\bibfnamefont {L.~J.}\ \bibnamefont {Gauckler}},\ }\href@noop {}
  {\bibfield  {journal} {\bibinfo  {journal} {J. Coll. Interf. Sci.}\ }\textbf
  {\bibinfo {volume} {240}},\ \bibinfo {pages} {89} (\bibinfo {year}
  {2001})}\BibitemShut {NoStop}%
\bibitem [{\citenamefont {Nugent}\ \emph {et~al.}(2007)\citenamefont {Nugent},
  \citenamefont {Edmond}, \citenamefont {Patel},\ and\ \citenamefont
  {Weeks}}]{Weeks2007}%
  \BibitemOpen
  \bibfield  {author} {\bibinfo {author} {\bibfnamefont {C.~R.}\ \bibnamefont
  {Nugent}}, \bibinfo {author} {\bibfnamefont {K.~V.}\ \bibnamefont {Edmond}},
  \bibinfo {author} {\bibfnamefont {H.~N.}\ \bibnamefont {Patel}}, \ and\
  \bibinfo {author} {\bibfnamefont {E.~R.}\ \bibnamefont {Weeks}},\ }\href@noop
  {} {\bibfield  {journal} {\bibinfo  {journal} {Phys. Rev. Lett.}\ }\textbf
  {\bibinfo {volume} {99}},\ \bibinfo {pages} {025702} (\bibinfo {year}
  {2007})}\BibitemShut {NoStop}%
\bibitem [{\citenamefont {Ito}\ and\ \citenamefont
  {Hashimoto}(2006)}]{Ito2006}%
  \BibitemOpen
  \bibfield  {author} {\bibinfo {author} {\bibfnamefont {T.}~\bibnamefont
  {Ito}}\ and\ \bibinfo {author} {\bibfnamefont {Y.}~\bibnamefont
  {Hashimoto}},\ }\href@noop {} {\bibfield  {journal} {\bibinfo  {journal}
  {Journal of the Japanese and International Economies}\ }\textbf {\bibinfo
  {volume} {20}},\ \bibinfo {pages} {637} (\bibinfo {year} {2006})}\BibitemShut
  {NoStop}%
\bibitem [{\citenamefont {Malkiel}\ and\ \citenamefont
  {Fama}(1970)}]{Malkiel1973}%
  \BibitemOpen
  \bibfield  {author} {\bibinfo {author} {\bibfnamefont {B.~G.}\ \bibnamefont
  {Malkiel}}\ and\ \bibinfo {author} {\bibfnamefont {E.~F.}\ \bibnamefont
  {Fama}},\ }\href@noop {} {\bibfield  {journal} {\bibinfo  {journal} {J. of
  Fin.}\ }\textbf {\bibinfo {volume} {25}},\ \bibinfo {pages} {383} (\bibinfo
  {year} {1970})}\BibitemShut {NoStop}%
\bibitem [{\citenamefont {Peters}(1989)}]{Peters1989}%
  \BibitemOpen
  \bibfield  {author} {\bibinfo {author} {\bibfnamefont {E.~E.}\ \bibnamefont
  {Peters}},\ }\href@noop {} {\bibfield  {journal} {\bibinfo  {journal} {Fin.
  Anal. J.}\ }\textbf {\bibinfo {volume} {45}},\ \bibinfo {pages} {32}
  (\bibinfo {year} {1989})}\BibitemShut {NoStop}%
\end{thebibliography}
\end{document}


\title{Diffusive and arrested-like dynamics in currency exchange markets\\
  Supplemental Material}


\begin{abstract}
This supplemental material contains further results and comparisons with the theoretical model presented in the letter. The fittings of the pdfs and cdfs are presented in a format to allow a better comparison with the theory, and the price correlation functions, equivalent to the dynamic structure factor, and non-Gaussian parameters are presented for all time origins.
\end{abstract}
\date{\today}
\pacs{}
\maketitle

\begin{center}
{\bf{PRICE FLUCTUATION DISTRIBUTIONS}}
\end{center}

\begin{figure}[b]
  \includegraphics[scale=0.4]{SMF1.jpg}
  \caption{EURUSD pdfs for the periods $2010-2015$ (black circles), $2005-2009$ (green diamonds) and $2000-2004$ (orange squares) at $\tau = 5$min. (a), $\tau = 25$min. (b), $\tau = 125$min. (c), $\tau = 625$min. (d) and $\tau = 3125$min. (e). The blue line is the computation according to the slow dynamics pdf model, $G(p, t)$. The parameters employed in the theoretical line are the ones from the complementary cdf fit, i.e. $l=3.0 \times 10^{-3}$, $d=1.5 \times 10^{-3}$, $\tau_{1}=400$min. and $\tau_{2}=300$min.}
  \label{fig:pdf}
\end{figure}

Fig. 2(a) from the letter shows that different years resolve the same price probability distribution function (pdf), while the letter's Fig. 2(b) illustrates the fittings of the theoretical model to the experimental data at different lag time intervals, $\tau$. As information is strongly compressed in such figure, we replot the same data here, separating each value of $\tau$ in one panel as indicated in Fig. \ref{fig:pdf}. Here the pdf from different periods, $2000-2004$, $2005-2009$ and $2010-2015$ are plotted, and they all lay on the same master pdf, particularly, they all reproduce a double decay at large price differences. {\color{red}\st{Such finding is unexpected and hints that catastrophic events or fortunate ones have no statistical significance in currency markets, and it is tempting to relate such statement to currency markets being regulated.}}\\

\begin{figure}[b]
 \includegraphics[scale=0.395]{SMF2.jpg}
  \caption{EURUSD complementary cdfs for the period $2010 - 2015$, denoted by black squares. The corresponding fit according to the model proposed in our research and inspired by \cite{kob2007}, namely the compementary cdf to $G(p, t)$, is indicated by the blue line. Here fitting parameters are $l=3.0 \times 10^{-3}$, $d=1.5 \times 10^{-3}$, $\tau_{1}=400$min. and $\tau_{2}=300$min. The lag time in each figure is $\tau=5$min. (a), $\tau=25$min. (b), $\tau=125$min. (c), $\tau=625$min. (d) and $\tau=3125$min. (e).}
  \label{fig:cdf}
\end{figure}

In order to focus on the region of large displacements, where the apparent double decay in the pdf is observed it is more convenient to use the complementary cumulative distribution function (cdf), since the noise is reduced \cite{Clauset2009, Adler1998}. We have thus computed the complementary cdf to the $2010 - 2015$ experimental EURUSD fluctuation data, Fig. \ref{fig:cdf}. This allows us to resolve {\color{red}\st{unambiguously}} the double decay in the corresponding pdf, {\color{red}in striking contrast with other models \st{as well as to rule out a power law decay predicted by other models}} \cite{Clauset2009}. Additionally, this representation is helpful to assess the quality of the fitting in the region of large displacements. Remarkable agreement is found between the theoretical curves and the experimental EURUSD data. Considering that complementary cdfs can provide a more accurate estimate of the fitting parameters, we have in fact fitted the theoretical cdf to the experimental data, and employed the obtained fitting parameters in order to calculate the corresponding pdf as shown in Fig. \ref{fig:pdf}. 

The fitting of the theoretical model to the EURUSD data is very good, except at low $\tau$. Note that the values of parameters indicate that there is no separation between short time and long time dynamics, because $l>d$ and $\tau_1 \approx \tau_2$. This would correspond to a fluid above the glass temperature, and only slightly undercooled. 

\begin{center}
{\bf{CALCULATION OF THE THEORETICAL MSPD}}
\end{center}

The mean squared price displacement (MSPD) can be calculated from the pdf, $G(p,\tau)$, as:

\begin{equation}
\langle \Delta p^2(\tau) \rangle = \langle [p(t_0+\tau)-p(t_0)]^2\rangle =
\int_{-\infty}^{\infty} p^2 G(p,\tau) dp
\end{equation}

\noindent We have fitted the model presented in the letter to the daily EURUSD MSPDs from $2015$ taking $t_{0}$=09:30h and $t_{0}$=18:00h. Illustrated in Fig. \ref{fig:MSPDs-fit}, it can be observed that the model correctly reproduces the experimental data for both, the fast and slow dynamics (i.e. at commencing times $t_0$=9:30h and $t_0$=18:00h, respectively). The set of parameters obtained from the fittings are indicated in Table I. {\color{red}Please note that the computed MSPD is the second moment from the price fluctuation pdf $G(p, t)$. Here, it is worthwhile pointing out that the second moment to the Ornstein-Uhlenbeck process depicting the transient price trapping distribution, $f_{vib}(p, t)$ as considered in $G(p, t)$, is $\sigma_{OU}=l^{2}\{1-\exp(-2\alpha t)\}$.}\\

\begin{table}[b]
\caption{\label{tab:table1} Parameters employed to fit the experimental MSPDs at $24$h periods, for different commencing times.}
\begin{center}
\begin{tabular}{c|c|c|c|c|c}
Commencing time & $D$ & $l$  & $d$ & $\tau_1$ & $\tau_2$\\
$t_{0}$  & $ \left(\mbox{min.}^{-1}\right)$ &$\times 10^{3}$ & $\times 10^{3}$& (min.) & (min)\\ \hline
9:30h & $2\cdot 10^{-4}$ & $3.00$ & $1.50$ & 400 & 300 \\
18:00h & {\color{red}$10^{-9}$} & $0.10$ & $0.15$ & 400 & 300
\end{tabular}
\end{center}
\end{table}

In the fittings, $\tau_1$ and $\tau_2$ are forced to be constant for different $t_0$ and equal to the values obtained from the average distributions, Fig. 3 in the letter, in order to minimize the number of free parameters while preserving their physical meaning. {\color{red}Also, while $D=2\cdot 10^{-4}\,\mbox{min.}^{-1}$ at $t_{0}=$ 9:30h, it is set to $D=10^{-9}\,\mbox{min.}^{-1}$ at $t_{0}=$ 18:00h as the steady state is reached at $\alpha^{-1}\sim10\,\mbox{min.}$} Note that the values of $l$ and $d$ bracket the average values given by the fitting to the overall distributions and MSPD. For $t_0$=18:00h, where the short time and long time dynamics are separated, $l<d$, what implies a transient localization of $\langle \Delta p^2 \rangle$ around $l^2$. The fitting and physical interpretation of the parameters shows that our model can be used to rationalize experimental price fluctuations and MSPD.

\vspace{0.5cm}
\begin{figure}[t]
  \includegraphics[scale=0.45]{SMF3.jpg}
  \caption{Red squares denote the MSPD at $t_{0}$=18:00h while black circles belong to the MSPD at $t_{0}$=09:30h. The red and black lines are the corresponding fits according to the second moment of the fluctuation pdf $G(p, t)$.}
  \label{fig:MSPDs-fit}
\end{figure}

\begin{center}
{\bf{DYNAMIC STRUCTURE FACTOR $S(q, \tau)$}}
\end{center}

Experiments on colloidal glasses typically rely on the measurement of the intermediate scattering function, or dynamic structure factor, $S(q,\tau)$, usually via light scattering techniques. This function is the Fourier transform of the MSD, and gives the coherent relaxation of the system in a length-scale set by the wave-vector $q$. $S(q,t)$ decays from $1$ at $\tau=0$ to zero in a fluid or a finite value in a glass (known as non-ergodicity or Debye-Waller factor), while for undercooled fluids shows a two step decay, with an intermediate plateau at the value of the glass non-ergodicity factor \cite{Binder05}. We have thus computed the analogue of the dynamic structure factor for the EURUSD, from the 2015 EURUSD $24$h datasets. Note however, that such calculation is not straightforward as we have not yet established an analogue to the Fourier reciprocal length depicted by the wave-vector $q$. However, we consider the localization distance, $r_{l}$ from the MSPD at $t_{0}$=18:00h, i.e. we consider that we sample the dynamics at length-scales of the order of $r_{l}$. Based on the hard spheres system, the two-step decay is nicely observed at $q \sim 2\pi/\sigma$, being $\sigma$ the colloidal diameter, whereas the localization length  $r_{l} \sim 0.1 \sigma$ \cite{Binder05}. Thus, we calculate the wave-vector as: 
\begin{equation}
q \sim 2\pi/\sigma = 2\pi/10r_{l}
\end{equation}
\noindent{\color{red}\st{where the additional $\sqrt{3}$ factor comes from the dimension of space.}} This yields {\color{red}\st{$q \sim 6.8\times10^3$}}$q \sim 3.9\times10^3$ . The dynamic structure factor, or price correlation function, is then calculated as:
\begin{equation}
S(q, \tau) = Re\left<\exp\{iq(p(t_{0}+\tau)-p(t_{0}))\}\right>
\end{equation}
\noindent in symmetry with the colloidal field, where vectorial quantities are used \cite{AFN-AMP2016}. The EURUSD $S(q, \tau)$ computed at different commencing times are displayed in Fig. \ref{Sqtau}. While the correlation function decays smoothly to zero for $t_0$=09:30h, it shows a two-step decay for $t_0$=18:00h, with a gradual evolution between these commencing times, Fig.\ref{Sqtau}(a). The transition is reversed when $t_0$ is further increased, Fig.\ref{Sqtau}(b). The behaviour found at $t_0$=18:00h is similar to that typical of undercooled systems. Furthermore, the height of the intermediate plateau is at a magnitude of about 0.8, close to 1, where 1 belongs to total dynamic arrest, supporting our choice of the value of $q$. This graph shows that the dynamics of the EURUSD system evolves from high temperature fluid-like dynamics to an undercooled fluid-like one with $t_0$, in agreement with the conclusion extracted from the MSPD displayed in the letter, Fig. 4. 

\begin{figure}[t]
  \includegraphics[scale=0.65]{SMF4.jpg}
  \caption{2015 EURUSD Dynamic structure factors $S(q, \tau)$ when considering $24$h periods, computed at different reference times $t_{0}$. The range $t_{0}$=09:30h - 18:00h (a), indicates an onset towards arrested dynamics as $S(q, \tau)$ gradually exhibit a slower decay when $t_{0}$ is raised from 09:30h, and eventually display a double decay featured by a plateau for $t_{0}$=18:00h. The opposite scenario is observed in the range $t_{0}$=18:00h - 09:30h, as $S(q, \tau)$ gradually shift to a profile characterized by a single decay while the intermediate plateau vanishes (b). The lines are fits to the data as indicated in the text. The black line is the fit at $t_{0}$=09:30h and the red line is the fit at $t_{0}$=18:00h. In each graph, legends indicate the corresponding $t_{0}$ time of each dataset.
  \label{Sqtau}}
\end{figure}

We have characterized $S(q, \tau)$ by fitting the data at $t_{0}$=09:30h to a single exponential functional form, $S(q, \tau) = \exp{(-(\tau/\tau_{0})^{p})}$, typical of high temperature fluids, and find $p$=0.95, $\tau$=3.25min. Then, $S(q, \tau)$ at $t_{0}$=18:00h is adjusted to a double exponential form,  $S(q, \tau) = A_{1}\exp{(-(\tau/\tau_{\alpha})^{\alpha})}+(1-A_{1})\exp{(-(\tau/\tau_{\beta})^{\beta})}$. We found $\tau_{\alpha}$=0.75min., $\alpha$=1.05, $\tau_{\beta}$=60min., $\beta$=1.4 and the amplitude $A_{1}$=0.15. Note that while at $t_{0}$=09:30h $S(q, \tau)$ is described by diffusive dynamics, at $t_{0}$=18:00h, the magnitudes of exponents $\alpha$ and $\beta$ are the ones typically found in arrested systems, in particular in undercooled and glassy colloids. Detailed inspection of the parameters indicate that the final relaxation corresponds to a compressed exponential (instead of a stretched exponential). Compressed exponentials in the decay of the intermediate scattering function have been previously found in colloidal gels \cite{Cipelletti2000}, focusing the symmetry with colloidal gels.

\begin{center}
{\bf{NON-GAUSSIAN PARAMETER, $\alpha_{2}(\tau)$}}
\end{center}

One further parameter that is usually monitored in physical glasses is the non-Gaussian parameter, $\alpha_2$, which gives a measure of the deviation of the pdf from the Gaussian shape. $\alpha_2$ is calculated as the ratio of the fourth to the second moments of the distribution, with the corresponding magnitude for a Gaussian distribution subtracted \cite{Rahman1964, Kob1997, Weeks2002}. In our one-dimensional problem, the definition of $\alpha_2$:

\begin{equation}
\alpha_{2} (\tau) = \frac{\left< \Delta p\big(\tau\big)^{4} \right>}{3\left< \Delta p\big(\tau\big)^{2} \right>^{2}} -1
\end{equation}

\noindent is slightly varied from the conventional form of three-dimensional glasses,
where $\Delta p(\tau) = p(t_{0}+\tau)-p(t_{0})$ is the relative difference of the currency pair price at time $t$, $p(t)$, and $t+\tau$. By definition, $\alpha_2$ is zero for Brownian dynamics, where displacement distributions are Gaussian, while its magnitude arises due to hallmarks from slow dynamics such as dynamical heterogeneities or anisotropies. Thus, the increase of $\alpha_{2}(\tau)$ is more prominent at lag-times $\tau$, where slow relaxations take place, which yields a bell-like profile to non-Gaussian parameters corresponding to undercooled fluids (at long times, Brownian diffusion is recovered). Therefore, $\alpha_{2}(\tau)$ is an excellent tool when identifying diffusive, undercooled and arrested dynamics. 

For the EURUSD system, we have computed $\alpha_{2}(\tau)$ to $2015$ daily datasets, at different reference times expecting to find different behaviors for different $t_{0}$. The non-Gaussian parameter indeed displays a gradually increasing maximum and a broader profile as the reference time of 18:00h is reached from the initial time of $t_{0}$=09:30h, depicted in Fig. \ref{alpha2}(a). Note that at $t_{0}$=18:00h the maximum is highest and the non-Gaussian parameter is broadest, in the time range when the plateaus in both the MSPD and $S(q,t)$ are found. When the reference time is increased further to 09:30h, the peak and $\alpha_{2}$'s profile gradually vanish at intermediate $t_{0}$, with $\alpha_{2}(\tau)$ being at the noise level at $t_{0}$=09:30h as observed in \ref{alpha2}(b). Therefore, the rise of a maximum in the non-Gaussian parameter indicates a gradual dynamic slowing down when less active market periods are reached, in agreement with the MSPD and dynamic structure factors presented above. 
\begin{figure}[t]
  \includegraphics[scale=0.65]{SMF5.jpg}
  \caption{2015 EURUSD non-gaussian parameter, $\alpha_{2}(\tau)$ at time intervals of $24$h and different commencing times $t_{0}$ as indicated in each figure. Note the peaked profile at $t_{0}$=18:00h, while at $t_{0}$=09:30h $\alpha_{2}(\tau)$ is at residual noise.
  \label{alpha2}}
\end{figure}
The different behaviors found for the EURUSD dynamics can be identified as different states transiting from diffusion to undercooled or arrested dynamics in systems such as highly correlated liquids or glasses. The theoretical model presented above can reproduce quantitatively these different behaviors, supporting the symmetry between these two far-apart fields, namely, foreign exchange market and the dynamics of undercooled fluids.


%